\documentstyle[aps,preprint,epsf]{revtex}

\newcommand{\be}[1]{\begin{equation}\label{#1}}
\newcommand{\ee}{\end{equation}}

\newcommand{\bea}[1]{\begin{eqnarray}\label{#1}}
\newcommand{\eea}{\end{eqnarray}}
\newcommand{\ra}{\rightarrow}

\newcommand{\prlput}[1]{}
\newcommand{\rem}[1]{}

\newcommand{\eref}[1]{(\ref{#1})}
\newcommand{\Eref}[1]{Eq.~(\ref{#1})}

\newcommand{\NN}{{\cal N}}

\def\eps{\epsilon}

\def\susy{supersymmetry}
\newcommand{\vev}[1]{\langle#1\rangle}
\def\susic{supersymmetric}

\def\Lgr{{\cal L}}
\def\QQ{{\cal Q}}
\def\KK{{\cal K}}
\def\DD{{\cal D}}
\def\VV{{\cal V}}
\def\NN{{\cal N}}
\def\UU{{\cal U}}

\def\none{$\NN=1$}
\def\ntwo{$\NN=2$}

\def\nthree{$\NN=3$}
\def\nfour{$\NN=4$}

\def\susy{supersymmetry}
\def\susic{supersymmetric}

\def\rarr{\rightarrow}

\def\ta{\theta}

\def\Qt{{\tilde Q}}
\def\mn{{\mu\nu}}
\def\del{\partial}

\begin{document}
%\draft
%\pagestyle{empty}
{\tightenlines

\preprint{
\begin{minipage}[t]{3in}
\begin{flushright}
IASSNS--HEP--99/15\\
hep-th/9902033 \\
\end{flushright}
\end{minipage}
}

\title{\Large\bf On Mirror Symmetry \\ in \\ Three Dimensional
Abelian Gauge Theories}
\author{{\bf
Anton Kapustin} and
{\bf
Matthew J. Strassler}}
\address{\ \\ School of Natural Sciences,
Institute for Advanced Study,
Olden Lane,
Princeton, NJ 08540, USA\\ ~ \\
{\tt kapustin, strasslr@ias.edu}}
\maketitle
\begin{abstract}
We present an identity relating the partition function of \nfour\
supersymmetric QED to that of its dual under mirror symmetry.  The
identity is a generalized Fourier transform.  Many known properties of
abelian theories can be derived from this formula, including the
mirror transforms for more general gauge and matter content.  We show
that \nthree\ Chern-Simons QED and \nfour\ QED with BF-type couplings
are conformal field theories with exactly marginal couplings.  Mirror
symmetry acts on these theories as strong-weak coupling duality.
After identifying the mirror of the gauge coupling (sometimes called
the ``magnetic coupling'') we construct a theory which is {\it
exactly} mirror --- at all scales --- to \nfour\ SQED.  We also study
vortex-creation operators in the large $N_f$ limit.

\end{abstract}

%\pacs{???}

\newpage
\pagestyle{plain}
\narrowtext
%\widetext

\vskip 0.4 in

\draft
\tightenlines

\section{Introduction}

Major advances in supersymmetric field theory and string theory in
various dimensions have led to the understanding that it is common for
apparently different quantum field theories to be quantum-mechanically
equivalent. Two theories which are ``dual'' in this way may be thought
of as two choices of variables in a path integral representation for
the same generating functional. Not that such ``duality relations''
are new; the relation between position space and momentum space
representations of quantum mechanical systems are of this type; the
order-disorder-fermion representations of the Ising model, the
identity of the sine-Gordon model and the Thirring model,
and target-space duality in two-dimensional sigma-models
are well known from two dimensions; and it has long
been conjectured that \nfour\ \susic\ Yang-Mills theory in four
dimensions is a conformal field theory with a duality symmetry.  The
developments in the last few years have provided vast amounts of
circumstantial evidence for the latter conjecture and have shown that 
many
different duality transformations exist in higher dimensions with as
few as four supercharges (which is \none\ \susy\ in four dimensions
and \ntwo\ \susy\ in three.)

However, outside of a small number of examples --- free field
theories, some lattice models, and a few two-dimensional continuum
field theories --- we do not know the precise change of variables
which would allow the transformation from one representation of a
theory to a dual representation.  In this paper we take a small step
toward making the ``mirror symmetry''\cite{kinsddd} of three
dimensional \nfour\ \susic\ abelian gauge theories explicit.  First,
focusing on the infrared behavior of these theories, where mirror
symmetry is exact, we present a formula which captures the
essence of the mirror symmetry transformation in the form of a
generalized Fourier transform. This formula encodes
most known results in abelian mirror symmetry in simple ways. Second,
we use the formula to derive some new results.  We consider \nthree\
Chern-Simons theories \cite{zupnik,kaoleelee} and \nfour\ theories
with BF-type couplings \cite{BrooksGates} interacting with matter, and
argue they flow in the infrared to lines of fixed points parameterized
by the coefficient of the CS or BF term. As we will show, mirror
symmetry maps these models to models of the same type while inverting
the CS or BF coupling; the inversion of the CS coupling agrees with
\cite{kooCS}.  Third, after identifying the field theory origin of the
mirror of the gauge coupling (the so-called ``magnetic coupling''), we
use our formula to suggest a mirror for \nfour\ SQED which is valid at
all energy scales, not just in the infrared.   Finally, we
discuss the construction of the vortex-creation operators in
\nfour\ SQED, and compute their dimension at large
$N_f$.

\section{Preliminaries}

We work in Minkowski space with signature $(-++)$. The \nfour\
superalgebra has eight supercharges which are doublets under
$SL(2,{\bf R})\times SU(2)_R\times SU(2)_N$; the first factor is the
Lorenz group, while the last two are R-symmetries.  We use indices
$\alpha,\beta; i,j; a,b$ for the indices of the defining
representation of these three factors.  The abelian gauge theories
which are the subject of this paper describe the interaction of $U(1)$
vector multiplets $\VV$ and charged hypermultiplets $\QQ$.  In
components the vector multiplet contains a gauge boson $A_\mu$, a
gaugino $\lambda_{a\alpha }^i$ and three real scalars $\Phi^{\{ij\}}$,
while the hypermultiplet contains a doublet of complex scalars $Q_a$
and a doublet of spinors $\psi_{i\alpha}$.  Because \nfour\ superspace
is often inconvenient, we will use \ntwo\ superspace language. The
hypermultiplet can be written as two \ntwo\ chiral superfields $Q,\Qt$
of charge $1,-1$.  The \nfour\ vector multiplet consists of an \ntwo\
real vector multiplet $V$ whose lowest component is a real scalar and
a chiral multiplet $\Phi$ whose lowest component is a complex
scalar. In the \ntwo\ notation only the $U(1)_N\subset SU(2)_N$
R-symmetry is explicit; the superfield $\Phi$ has $U(1)_N$ charge $2$,
while the rest of the \ntwo\ superfields are uncharged.

The \nfour\ supersymmetry algebra has an idempotent outer automorphism
which interchanges $SU(2)_R$ and $SU(2)_N$. This automorphism takes an
ordinary vector multiplet, whose scalars transform as a triplet of
$SU(2)_N$, into a twisted vector multiplet~\cite{BrooksGates}, whose
scalars transform as a triplet of $SU(2)_R$. Similarly, one can define
a twisted hypermultiplet whose bosonic fields form an $SU(2)_N$
doublet.  Field and superfield constituents of twisted \nfour\
multiplets will be distinguished with a hat, e.g. $\hat\Phi$ for the
chiral part of the twisted vector multiplet. Note that $\hat\Phi$ has
$U(1)_N$ charge $0$, while $\hat{Q}$ and $\hat{\Qt}$, the chiral
constituents of the twisted hypermultiplet, have $U(1)_N$ charge $1$.

In three dimensions a photon is the electric-magnetic dual of a
scalar.  The scalar is periodic if the gauge group is compact, and
shifting it by a constant is a symmetry of the classical theory.  This
duality transformation takes a free \nfour\ vector multiplet into a
twisted hypermultiplet with target space ${\bf R}^3\times {\bf S}^1$
or ${\bf R}^4$, depending on whether the gauge group is compact or
not. Similarly, a free twisted vector multiplet is the electric-magnetic
dual of an ordinary hypermultiplet.

Yet another type of \nfour\ multiplet is a linear multiplet (which
also has a twisted version). Linear multiplets are important because
they contain conserved currents. An \nfour\ linear multiplet consists
of an \ntwo\ linear multiplet $\Sigma$ (a real superfield satisfying
$D^2\Sigma=\bar D^2\Sigma=0$) and an \ntwo\ chiral superfield $\Pi$
with $U(1)_N$ charge $2$. The field strength of a vector multiplet $F$
resides in a linear multiplet with $\Sigma=iD\bar D V$ and
$\Pi=\Phi$. The conserved current in this case is $^*F$ where $*$ is
the Hodge star; the conservation of $^*F$ is a consequence of the
Bianchi identity $dF=0$. The charge associated to this current is the
generator of the shift symmetry of the dual photon mentioned above.
The Noether current associated with the flavor symmetries of a twisted
hypermultiplet also resides in an \nfour\ linear multiplet; in this
case $\Sigma=\hat{Q}^\dagger\hat{Q}- \hat{\Qt}\hat{\Qt}^\dagger,\
\Pi=\hat{\Qt} \hat{Q}$.  Conversely, the topological current
$^*\hat{F}$ associated with a twisted vector multiplet and the flavor
current of an ordinary hypermultiplet reside in a twisted linear
multiplet. The latter consists of an \ntwo\ linear multiplet
$\hat\Sigma$ and a chiral multiplet $\hat\Pi$ with $U(1)_N$ charge
$0$.

The action of an \nfour\ theory of hypermultiplets and abelian vector
multiplets contains kinetic terms for the hypermultiplets
\be{SQEDa}
S_{H}(\QQ,\VV) = -\int d^3x\ d^4\ta\ 
\left(Q^\dag e^{2V} Q + \Qt^\dag e^{-2V}\Qt\right) 
- \left[\int d^3x\ d^2\ta\ i{\sqrt 2}\Phi Q\Qt + 
c.c.\right],\ 
\ee
which include scalar kinetic terms $-|D_\mu Q|^2-|D_\mu \Qt|^2$,
and kinetic terms for the vector multiplets 
\be{SQEDb}
{1\over g^2}S_{V}(\VV) ={1\over g^2}\int d^3x\ d^4\ta\ 
\left\{
{1\over 4}\Sigma^2 - \Phi^\dag \Phi \right\} \ 
\ee
which include ${1\over 4g^2}F_{\mu\nu}^2$. One can also add 
mass terms for the  hypermultiplets
\be{SQEDmass}
S_m(\QQ)=-\int d^3x\ d^4\ta
\left(Q^\dag e^{-2im_r\ta\bar{\ta}} Q + \Qt^\dag e^{
2im_r\ta\bar{\ta}}\Qt\right)
-\left[\int d^3x\ d^2\ta\ m Q\Qt +c.c.\right]
\ee
and Fayet-Iliopoulos (FI) terms for the vector multiplets
\be{SQEDFI}
S_{FI}(\VV)={\xi_r\over \pi}\int d^3x\ d^4\ta\ V-\left[
\frac{i\xi}{2\pi}\int d^3x\ d^2\ta\ 
\Phi +c.c\right].
\ee
Here $m_r\in {\bf R}$ and $m\in {\bf C}$ together form an $SU(2)_N$
triplet while $\xi_r\in {\bf R}$ and $\xi\in {\bf C}$ form an
$SU(2)_R$ triplet.  These building blocks suffice to construct the
most general renormalizable action containing only ordinary
hypermultiplets and vector multiplets.  The most general
renormalizable action for twisted fields is obtained by putting hats
over all fields in \eref{SQEDa}--\eref{SQEDFI}.

If the mass terms and the FI terms are zero, the \nfour\ action has
$SU(2)_R\times SU(2)_N$ R-symmetry, as well as two discrete symmetries
which we call P and CP.  To define these discrete symmetries we need
to recall how parity transformation acts on Majorana spinors in three
dimensions. Let us define parity as a reflection of one of the spatial
coordinates, say $x^1\ra {x'}^1=-x^1$. To ensure parity-invariance of
the Dirac equation we must transform spinors according to $\psi\ra
R\psi$, where the two-by-two matrix $R$ satisfies $R^T R=1, \
R^T\gamma^0\gamma^1 R=-\gamma^0\gamma^1,\ R^T\gamma^0\gamma^2
R=\gamma^0\gamma^2$.  Then parity acts on \ntwo\ superspace via
${x'}^0=x^0,{x'}^1=-x^1,{x'}^2=x^2, \ta'=R\ta$. The chiral superspace
measure $d^2\ta$ is parity-odd, while $d^4\ta$ is parity-even. We
define P as a transformation which acts on \ntwo\ superfields via
\bea{P} 
V'(x',\ta')=V(x,\ta),\quad \Phi'(x',\ta')=-\Phi(x,\ta),\\
\nonumber Q'(x',\ta')=Q(x,\ta),\quad \Qt'(x',\ta')=\Qt(x,\ta).  
\eea
The transformation CP is defined by 
\bea{CP}
V'(x',\ta')=-V(x,\ta),\quad \Phi'(x',\ta')=\Phi(x,\ta),\\ \nonumber
Q'(x',\ta')=\Qt(x,\ta),\quad \Qt'(x',\ta')=-Q(x,\ta).  
\eea 
We call these transformations P and CP because the gauge field behaves
as a polar vector with respect to P and as an axial vector with
respect to CP. It easy to check that when masses and FI terms are
absent, the action is both P and CP-invariant.  Mass terms break P,
while FI terms break CP. To define P and CP for twisted multiplets we
simply put hats over all fields in Eqs.~\eref{P} and \eref{CP}.

Let us recall how mirror symmetry works for \nfour\ abelian gauge
theories without twisted fields~\cite{kinsddd}.  If we set the
hypermultiplet masses and FI couplings to zero, then the only mass
scale in these theories is $g^2$, and they are believed to flow to
nontrivial superconformal fixed points in the infrared, where the
scale $g^2$ is washed out.  Each of these superconformal fixed points
has a dual description using the duality mapping known as mirror
symmetry~\cite{kinsddd}.  Under mirror symmetry, electrically charged
particles and Abrikosov vortex solitons are exchanged.  The mirror
theory is a twisted abelian gauge theory, i.e.  the fundamental
degrees of freedom live in twisted hypermultiplets and twisted vector
multiplets.  The Higgs branch of one theory is the Coulomb branch of
its mirror; similarly, the mass terms for the hypermultiplets are
mirror to the FI terms for the twisted vector multiplets
(which determine the masses of vortices.)  The mapping of flavor
symmetries is generally complicated.  The $U(1)$ currents from abelian
subgroups of hypermultiplet flavor symmetries are mapped to the $U(1)$
currents $^*\hat{F}$ \cite{kinsddd}.  The off-diagonal currents of the
flavor symmetries are not seen semiclassically and will not be
discussed below.

For example, the mirror of \nfour\ $U(1)$ with $N_f$ flavors [we will
refer to this theory as SQED-$N_f$] is a twisted $U(1)^{N_f-1}$ gauge
theory with $N_f$ twisted hypermultiplets $\hat{Q}_p,\hat{\Qt}_p,
p=1,\ldots,N_f$, where $\hat{Q}_p$ has charge $+1$ under the $p^{th}$
$U(1)$ factor and charge $-1$ under the $(p-1)^{th}$ $U(1)$ factor
\cite{kinsddd}. The topological current $^*F$ of \nfour\ SQED is
mirror to the Noether current which generates a $U(1)$ global symmetry
transformation $\hat{Q}_p\ra e^{i\alpha} \hat{Q}_p,\hat{\Qt}_p\ra
e^{-i\alpha} \hat{\Qt}_p, p=1,\ldots,N_f$. It is convenient to choose
the normalization in which $\hat{Q}$ has charge $1/N_f$ under this
global $U(1)$; then all gauge invariant operators in the
$U(1)^{N_f-1}$ theory have integer global $U(1)$ charges.  Note that
the case $N_f=2$ is special, since the mirror theory is isomorphic to
the original one \cite{kinsddd}.

It is also interesting to consider theories which contain both
ordinary and twisted \nfour\ multiplets.  A natural way to couple
twisted and ordinary vector multiplets is by means of an \nfour\ BF
term~\cite{BrooksGates}. It appears that this is the only way to
couple twisted and ordinary fields without introducing operators of
dimension higher than three. The \nfour\ BF term has the following
form:
\be{SQEDbf} S_{BF}(\hat\VV,\VV) = {1\over 2\pi}\int d^3x\
d^4\ta\ V \hat\Sigma-\left[{1\over 2\pi}\int d^3x\ d^2\ta\ i\Phi
\hat\Phi + c.c.\right] \ee 
Its component form in the Wess-Zumino gauge is
given by 
\be{bfcomp} S_{BF}={1\over 2\pi}\left( -{1\over 2} \eps^{mnp}
A_m \hat{F}_{np}+ \Phi^{\{ij\}} \hat{D}_{\{ij\}}+\hat{\Phi}^{\{ab\}}
D_{\{ab\}}+ i\lambda_{a\alpha}^i\hat{\lambda}^{a\alpha}_i\right) \ .
\ee
Here $D_{\{ab\}}$ and $\hat{D}_{\{ij\}}$ are the auxiliary fields of
the ordinary and twisted vector multiplets respectively.  The authors
of \cite{BrooksGates}, who were the first to construct the \nfour\ BF
coupling, observed that both ordinary and twisted multiplets were
required.  Noting the analogy with two dimensions, they correctly
conjectured the existence of a mirror symmetry which would exchange
these multiplets.  We will see in Section III that the BF interaction
lies at the heart of the mirror transform.
 
The BF term gives a gauge-invariant mass to both $\VV$ and $\hat{\VV}$.
It also breaks P and CP. Certain discrete symmetries remain unbroken, 
however. Namely, a transformation which acts as P (CP) on the ordinary
fields and as CP (P) on the twisted fields is still a symmetry.

When a BF term is present in the action, one can dualize either a
twisted vector multiplet, or an ordinary one, but not both of them
simultaneously. Also, in the presence of the BF term the shift
symmetry of the dual photon is gauged. This will be discussed in more
detail in Section IV. 

We will also need gauge-fixing terms. Their explicit form is unimportant
for our purposes. For example, one can use an \ntwo--covariant
version of Landau gauge:
\be{SQEDd}
S_{GF}(\VV) = \int d^3x\left( \int d^2\ta\ 
\Psi {\bar D}^2 V+c.c. \right)\ ,
\ee
where $\Psi$ is a chiral superfield serving as a Lagrange multiplier.
This particular gauge-fixing term breaks \nfour\ supersymmetry down to
\ntwo\, but the correlators of gauge-invariant quantities remain 
\nfour--supersymmetric.

As in \cite{kinsddd}, one can prove nonrenormalization theorems for
various branches of the moduli space. In particular, the metric on the
Higgs branch, where ordinary hypermultiplets have VEVs, does not
depend on the gauge coupling of the ordinary vector multiplets.
Similarly, the metric on the twisted Higgs branch, where twisted
hypermultiplets have VEVs, is unaffected by the twisted gauge
coupling.  On the other hand, in the presence of the BF term the
metric on the Higgs branch does depend on the twisted gauge coupling.
We will see this explicitly in Section IV.

\section{The Mirror Transform is a Fourier Transform}

It has been known for some time that most known results of abelian
mirror symmetry can be derived from the properties of \nfour\
supersymmetric $U(1)$ gauge theory with a single charged
hypermultiplet (an electron, a positron and their scalar partners.)
This theory, which we will call SQED-1, flows from weak coupling in
the ultraviolet to strong coupling in the infrared, where it becomes a
conformal field theory (CFT) which we will refer to as CFT-1.  The
fundamental result of mirror symmetry is that CFT-1 is equivalent to a
Gaussian theory \cite{nsewddd} --- namely, a free twisted
hypermultiplet.

SQED-1 has a single abelian global symmetry whose current is
$^*F$. The associated charge, the integrated magnetic flux, is the
vortex number. We may couple this current and its superpartners to a
background twisted vector multiplet $\hat \VV$ through a BF-type
interaction \eref{SQEDbf}.  Then the generating functional for
correlation functions of the current multiplet in SQED-1 can be
written as
\be{SQEDPI} Z_{{\rm SQED-1}}[\hat\VV] = 
\int\ \DD\VV\ \DD\QQ\ \exp\left({i\over
g^2}S_{V}(\VV)+iS_{GF}(\VV)+iS_{BF}(\hat\VV,\VV) 
+ iS_{H}(\QQ,\VV)  \right) \
.  
\ee 
We define this functional integral as the sum of its expansion in powers of 
$g^2$. Since the theory is abelian, there are no instanton corrections.
Power counting and symmetries imply that there are no divergences in
this expansion, so no counterterms are needed in \eref{SQEDPI}.  Since
$g$ is the only available scale, the perturbative expansion is
actually an expansion in powers of $g^2/p$ where $p$ is momentum.  To
obtain the infrared CFT, one needs to resum the perturbative series
and take the limit $g\rarr \infty$. Applying this limit formally to
\eref{SQEDPI} we obtain the expression
\be{CFTPI}
Z_{{\rm {CFT-1}}}[\hat\VV] = \int\ \DD\VV\ \DD\QQ\ 
e^{iS_{GF}(\VV)+iS_{BF}(\hat\VV,\VV)
+ iS_{H}(\QQ,\VV) }  \ .
\ee

As mentioned above, CFT-1 is equivalent to a theory of a free twisted
hypermultiplet $\hat\QQ$, with the field strength of $\VV$ being
mapped to the abelian $U(1)$ flavor current of the twisted
hypermultiplet.  The appropriate path integral is
\be{freePI} Z_{\hat\QQ}[\hat\VV] = \int\ \DD\hat\QQ\
e^{iS_{H}(\hat\QQ,\hat\VV)} \ .  
\ee 
The statement of mirror symmetry is 
$Z_{\hat\QQ}[\hat\VV]=Z_{{\rm {CFT-1}}}[\hat\VV]$.  The integrals over the
hypermultiplets are quadratic and give a superdeterminant of the
supersymmetric Laplacian  $\KK$ on flat $d=3$ \nfour\ superspace.
Using this, we may write the equivalence of these
two generating functionals in the following suggestive form:
\be{master}
{\rm Sdet}\left(\KK[\hat\VV]\right) = 
\int\ \DD\VV\ e^{iS_{GF}(\VV)}e^{iS_{BF}(\hat\VV,\VV)} \ 
{\rm Sdet}\left(\KK[\VV]\right)\ .
\ee
Mirror symmetry between \nfour\ SQED-1 and the theory of a free twisted
hypermultiplet is thus related to the invariance of the
superdeterminant under a Fourier transform with respect to the
background fields.  Note that this is highly non-trivial, as neither
superdeterminant is Gaussian.

The relation \eref{master} encapsulates many known properties of mirror
symmetry and allows them to be rederived using elementary 
manipulations. 
We list a few examples here.

\subsection{Inverse and repeated Fourier transforms}
To invert the functional Fourier transform (FFT) we multiply both
sides of \eref{master} by
$\exp[-iS_{BF}(\VV',\hat\VV)-iS_{GF}(\hat\VV)]$, where $\VV'$ is
another background vector multiplet, and integrate both sides of
\eref{master} over $\hat\VV$. The physical meaning of these
manipulations is that gauging a global symmetry in one theory
corresponds to removing a gauge symmetry (sometimes called
``ungauging'') using a BF coupling in its mirror. If instead we apply
the FFT to ${\rm Sdet}(K[\VV])$ twice, we get ${\rm Sdet}(K[-\VV])$,
implying that the fourth power of the FFT is the identity
transformation.  (This is also true for the ordinary Fourier transform
on the space of $C^\infty$ functions of rapid decrease.) In the
string-theoretic approach of \cite{ahew} to mirror symmetry, the
mirror transform is effected by a generator $S$ of $SL(2,Z)$ which
also satisfies $S^4=1$.

\subsection{Mapping of operators}
As discussed in Sec.~2, mirror symmetry maps hypermultiplet masses to
Fayet-Iliopoulos terms, hypermultiplet flavor currents to topological
currents, and the Higgs branch of moduli space to the Coulomb branch.
These mappings can be easily seen in \Eref{master}.  A free
hypermultiplet has a Higgs branch parameterized by
$\vev{\hat{Q}\hat{\Qt}}$ and
$\vev{\hat{Q}^\dagger\hat{Q}-\hat{\Qt}^\dagger\hat{\Qt}}$, while CFT-1
has a Coulomb branch with coordinates $\vev{\Phi}$, $\vev{\Sigma}$.
That these branches are exchanged is made clear by taking derivatives
of the two sides of \eref{master} with respect to $\hat\Phi$, $\hat
V$.  Similarly, a constant expectation value for $\hat\Phi$ gives a
mass to the hypermultiplet $\hat\QQ$ while inducing a Fayet-Iliopoulos
coupling for $\VV$.  The background gauge field $\hat A_\mu$ couples
to the topological current of CFT1 and to the flavor current of the
free hypermultiplet.

\subsection{The convolution theorem}
The inverse FFT and the convolution theorem may be applied to derive
mirror symmetry in all other abelian \nfour\ theories.  For example,
to study (twisted) SQED-$N_f$, with (twisted) hypermultiplets
$\hat{\QQ}_i$ of charge $1$, one raises both sides of \Eref{master} to
the power $N_f$ and then integrates over $\hat\VV$. On the left-hand
side one gets the partition function of the twisted SQED-$N_f$. On the
right-hand side, the integration over $\hat\VV$ removes the vector
multiplet which couples equally to all $N_f$ hypermultiplets.  The
remaining $N_f-1$ vectors and $N_f$ hypermultiplets form the
$U(1)^{N_f-1}$ theory described in Sec.~II.  Similar manipulations
allow one to find the mirror of an arbitrary abelian \nfour\ theory.
The results agree with \cite{ucbbrane}.

\subsection{\ntwo\ mirror symmetry}
 
\ntwo\ SQED with two oppositely charged chiral superfields $Q,\Qt$ can
be obtained from \nfour\ SQED-1 by coupling the latter to a neutral
chiral superfield $S$ via the interaction $\int d^2\ta\ S\Phi$, which
makes both $S$ and $\Phi$ massive \cite{ntwobrane,ntwovort}.  We may
identify the chiral field $\hat\Phi$ in the twisted vector multiplet
$\hat\VV$ with $S$ and integrate over $\hat\Phi$ on both sides of
\eref{master} with weight $\exp(-i/h \int d^4\ta\
\hat\Phi^\dagger\hat\Phi)$.  In our normalization $\hat\Phi$ has
engineering dimension $1$, therefore $h$ is a parameter of dimension
$1$. The right-hand side becomes the partition function of a theory
whose infrared (large $h$) limit is the same as the infrared limit of
\ntwo\ SQED-1.  The left-hand side is a partition function of an
\ntwo\ theory of three chiral superfields $\hat{Q},\hat{\Qt},\hat\Phi$
coupled via the superpotential $W=\hat\Phi\hat{Q}\hat{\Qt}$.  These
two theories were shown to be mirror in
\cite{ntwobrane,ntwovort}. Mirror symmetry in all other \ntwo\ abelian
theories can again be derived using the convolution theorem and the
inverse Fourier transform. In all these theories there are no
ultraviolet divergences (if regularization preserves all the
symmetries), so our formal manipulations are presumably justified.

\section{\nfour\ theories with BF couplings}

In this section we study \nfour\ theories which contain both ordinary
and twisted fields coupled via a BF term \Eref{SQEDbf}. These theories
apparently have not been considered in the literature. We will show
that BF couplings are exactly marginal.  They parameterize manifolds of
conformal field theories, in analogy to Maxwell couplings in finite
$d=4$ \ntwo\ supersymmetric gauge theories.  As in the
four-dimensional case, weakly coupled theories are found near certain
boundary points of these manifolds, with the inverse BF couplings
serving as expansion parameters for a finite perturbation series around a 
free theory.  Mirror symmetry acts on these manifolds by exchanging strongly 
coupled SCFTs with weakly coupled ones.

To be concrete, let us consider a copy of \nfour\ SQED-1 and 
a copy of its
twisted version, coupled via a BF-term with coefficient $k$.  
The classical action of this theory is
\be{BFaction}
 {1\over g^2} S_V(\VV) + {1\over \hat g^2} S_V(\hat\VV) +
S_H(\QQ,\VV)+S_H(\hat\QQ,\hat\VV)+kS_{BF}(\VV,\hat\VV)\ .
\ee
The vector multiplets become topologically massive and therefore both
the twisted and ordinary Coulomb branches are lifted.  The classical
moduli space of this theory consists of a Higgs branch and a twisted
Higgs branch parameterized by the ordinary and twisted hypermultiplet
vacuum expectation values, respectively.  These branches intersect at
a single point (the origin.)  To determine the metric we must solve
the D-flatness conditions modulo gauge transformations. The D-flatness
conditions for the Higgs branch are
\be{hyperK} \QQ^\dagger
\sigma^p \QQ=\frac{k}{2\pi}\hat\Phi^p\ , \ \ p=1,2,3.  
\ee 
Note that the lowest component of the twisted vector multiplet plays
the role of the dynamical Fayet-Iliopoulos term.  We expect that
Eqs.~\eref{hyperK} can be interpreted as moment map equations for a
hyperk\"ahler quotient~\cite{HKLR} (see also \cite{GR} for a review).

To see that this is indeed the case, recall that we can dualize the
twisted photon $\hat{A}$ into a scalar $\hat\tau$. $\hat\tau$ can be
combined with $\hat{\Phi}^p$ into a quaternion
$$w=\frac{\hat{g}^2\hat{\tau}}{\pi\sqrt{2}}+i\sigma^p\hat{\Phi}^p\ ,$$
where $\hat{g}$ is the twisted gauge coupling. In terms of $w$ the kinetic 
energy of the twisted vector multiplet takes the form 
$${1\over 2\hat{g}^2} |\partial w|^2\ .$$ 
The metric $|dw|^2/(2\hat{g}^2)$ is, up to an overall factor, the
standard hyperk\"ahler metric on ${\bf R}^4$ (or ${\bf R}^3\times {\bf
S}^1$ if we take the gauge group to be compact.) We can also think of
the complex doublet $\QQ$ as a quaternion, with metric $|d\QQ|^2$.
Under a constant gauge transformation from the untwisted $U(1)$, $\QQ$
transforms as $\QQ\ra \QQ e^{i\sigma_1 \alpha}$. Less trivially, $w$
transforms as $w\ra w-k \hat{g}^2 \alpha/(\pi\sqrt{2})$, as explained
below. The hyperk\"ahler moment equations for this transformation are
precisely \eref{hyperK}.  It is well known that this hyperk\"ahler
quotient yields the Taub-NUT metric~\cite{GR}, and thus the Higgs
branch is the Taub-NUT space.  This space can be thought of as a
circle fibered over ${\bf R}^3$. The Taub-NUT metric depends on a
single parameter which sets the asymptotic radius of this circle.  In
the present case this radius is $k\hat{g}$.  Identical arguments show
that the twisted Higgs branch is also a Taub-NUT space, the asymptotic
radius being $kg$. Note that these results are in agreement with the
nonrenormalization theorem stated in Section II, which says that the
metric on the Higgs branch (resp. twisted Higgs branch) does not
depend on the gauge coupling (resp. twisted gauge coupling).

Let us recall why the dual photon $\hat\tau$ transforms additively,
$\hat{\tau}\ra \hat{\tau}-k\alpha$, under a constant gauge
transformation. In short, the BF coupling $A\wedge \hat{F}$ can be
interpreted as a coupling of the gauge field $A$ to a topological
current $^*\hat{F}$ generating the shift of the dual photon. This
means that the shift symmetry is gauged, $A$ being the corresponding
gauge field, so a gauge transformation of $A$ must be accompanied by a
shift of $\hat{\tau}$. A more detailed argument goes as follows. In
order to dualize the twisted gauge field $\hat{A}$ we need to treat
its field strength $\hat{F}$ as an unconstrained 2-form and impose the
Bianchi identity $d\hat{F}=0$ via a Lagrange multiplier
$\hat\tau$. Then the action takes the form
$$-{1\over 4\pi} \eps^{mnp} \hat{F}_{mn} (k A_p+\partial_p
\hat{\tau})+\ldots\ ,$$ 
where dots denote terms which are manifestly
invariant with respect to gauge transformations $A\ra A+d\alpha$. The
action will be invariant if we also transform $\hat{\tau}$ as
$\hat{\tau}\ra\hat{\tau}-k\alpha$.

Our discussion of the metric was classical, but one can show that
there are no quantum-mechanical corrections. Indeed, supersymmetry
tells us that the metric is hyperk\"ahler, and we also know that it
has $SU(2)_R$ isometry ($SU(2)_N$ for the twisted Higgs branch) which
rotates the three complex structures. This, together with the known
asymptotic behavior, uniquely determines the metric~\cite{AtHi}.

In the infrared limit we must take both gauge couplings to infinity,
and then the moduli space becomes a pair of ${\bf R}^4$'s intersecting
at the origin.  For infinite gauge couplings the one-particle poles in
the propagators of the vector multiplets move to infinity;
nevertheless the vector multiplets cannot be ignored, since the BF
term remains and induces a nontrivial interaction between the ordinary
and twisted hypermultiplets, whose strength depends on $k$.  The
theory at the origin of the moduli space is a nontrivial CFT (as the
moduli space is not smooth there) with \nfour\ SUSY and unbroken
$SU(2)_R\times SU(2)_N$ symmetry.

For $k\ra\infty$ the vector multiplets decouple, so the CFT at the
origin becomes a direct sum of a free hypermultiplet and a free
twisted hypermultiplet. It is straightforward to set up perturbation
theory in $1/k$. Using the approach of
Refs.~\cite{csrenblasi,csrenkp2} it is easy to show that the
coefficient of the BF-term $k$ is not renormalized, so the CFT at the
origin is an exactly marginal deformation of the theory with
$k=\infty$, i.e. of a free theory. In fact, the high degree of
supersymmetry ensures that there are no ultraviolet divergences in
this expansion.  The dimension of any operator can be computed as a
power series in $1/k$.  The dimensions of operators in short
representations of the superconformal algebra are determined by their
$SU(2)_R\times SU(2)_N$ quantum numbers: $\Delta=j_R+j_N$ for scalar
primary operators, where $j_R$ and $j_N$ are the $SU(2)_R$ and
$SU(2)_N$ spins~\cite{Minwalla}.  The dimensions of operators in long
superconformal multiplets depend on $k$, in general.

In the opposite limit, $k\ra 0$, ordinary and twisted multiplets do
not couple to each other and the theory flows to a direct sum of CFT-1
and twisted CFT-1. Note that this theory is mirror to the theory at
$k\ra\infty$.  One may conjecture that more generally the CFT at large
$k$ and the CFT at small $k$ are mirror to each other. To show that
this is indeed the case, consider the following generating functional
for the CFT with BF coupling $k$:
\bea{BFone}
Z_k[\UU,\hat\UU]&=&\int \DD\VV\ \DD\hat{\VV}\ {\rm 
Sdet}\left(\KK[\hat\VV]\right) 
{\rm Sdet}\left(\KK[\VV]\right) 
\exp\left(ikS_{BF}(\hat{\VV},\VV)+iS_{GF}(\VV)+
iS_{GF}(\hat{\VV})+\right. \nonumber \\
& & \left. iS_{BF}(\hat\VV,\UU)+iS_{BF}(\hat\UU,\VV)\right)\ .
\eea
Differentiating $Z_k$ with respect to $\UU$ and $\hat\UU$ generates
all the correlations functions of ordinary and twisted vector
multiplets. Now we substitute the right-hand side of \eref{master} for
${\rm Sdet}\left(\KK[\VV]\right)$ and ${\rm
Sdet}\left(\KK[\hat\VV]\right)$ and perform Gaussian integrals over
$\VV$ and $\hat\VV$. The result turns out to be
\be{BFtwo}
Z_{-1/k}\left[-{1\over k}\UU,-{1\over k}\hat\UU\right]
\exp\left(-{i\over k} S_{BF}(\hat\UU,\UU)\right) \ .
\ee 
This means that the connected correlation functions of $\VV$ and
$\hat\VV$ in CFTs with BF couplings $k$ and $-1/k$ are related by a
trivial rescaling. The two-point functions in addition differ by a
contact term.

Our discussion can be easily generalized to theories with $N_f>1$ and
larger gauge groups.  For example, take $N_f$ copies of CFT-1, each
with a hypermultiplet of charge 1.  Take a similar set of copies of
twisted CFT-1, and couple the vector multiplets to the twisted vector
multiplets via a BF term, giving the action
$$ S_H(\QQ_i,\VV_i) + S_H(\hat\QQ_i,\hat\VV_i) 
 + \sum_{i,j} k_{ij} S_{BF}(\VV_i,\hat\VV_j)\ .$$ 
This yields a manifold of \nfour\ SCFTs parameterized by the matrix
$k$.  Mirror symmetry acts on this manifold by $k\ra -k^{-1}$.  When
$k$ is nondegenerate, the Coulomb branches are lifted.  The metrics of
the Higgs and twisted Higgs branches can be computed as in the $N_f=1$
case above and, for finite Maxwell couplings, turn out to be of the
Lindstr\"om-Ro\v{c}ek/Lee-Weinberg-Yi (LR/LWY) type~\cite{STdual,GR}.

In summary, \nfour\ theories with BF couplings are similar in many
respects to certain finite $N=2$ theories in four dimensions. Both
types of theories are finite in perturbation theory and have exactly
marginal couplings (real in $d=3$ and complex in $d=4$) which are
acted upon by a duality transformation. In $d=3$ this duality is
mirror symmetry, while in $d=4$ it is electric-magnetic duality.

\section{\nthree\ theories with Chern-Simons couplings}

It is believed impossible to write down an \nfour\ supersymmetric
Chern-Simons (CS) action coupled to matter. However, an \nthree\ CS
action exists~\cite{zupnik,kaoleelee}. One simply identifies the
ordinary and twisted vector multiplets appearing in the BF action
\eref{SQEDbf}. This identification obviously breaks $SU(2)_R\times
SU(2)_N$ symmetry down to its diagonal subgroup $SU(2)_D$ and, less
obviously, breaks \nfour\ SUSY down to \nthree.  Under $SU(2)_D$ the
four supercharges of \nfour\ decompose as ${\bf 1}+{\bf 3}$.  The
identification breaks the singlet while the triplet survives.  We will
see the properties of these theories are very similar to those
considered in the previous section.

Let us briefly review \nthree\ SUSY theories.  The basic multiplets
are the hypermultiplet and the vector multiplet.  The hypermultiplet
contains an $SU(2)_D$-doublet of scalars $Q_a$ and an
$SU(2)_D$-doublet of spinors $\psi_a$.  The vector multiplet contains
a triplet of real scalars $\Phi^{\{ab\}}$, a triplet of spinors
$\lambda^{\{ab\}}$, a singlet spinor $\lambda^0$, and a gauge boson
$A_\mu$.  An \nthree\ action for hypermultiplets automatically has
\nfour\ SUSY.  In particular an \nthree\ sigma-model must have a
hyperk\"ahler target space.  If restrict ourselves to renormalizable
theories, then the most general \nthree\ action for hypermultiplets
interacting with abelian vector multiplets is the sum of the \nfour\
action and the \nthree\ Chern-Simons term for the vector multiplets
\be{CS} 
\sum_{i,j} k_{ij} S_{CS}(\VV_i,\VV_j)=\sum_{i,j}
\frac{k_{ij}}{4\pi}\left(\int d^3 x d^4\ta\ \Sigma_i
V_j-\left[\int d^3x d^2\ta\ \Phi_i\Phi_j+c.c.\right]\right)\ .  
\ee 
Here $k_{ij}$ is a real symmetric matrix.\footnote{We take the gauge
group to be noncompact, i.e.  ${\bf R}^n$ rather than $U(1)^n$.  Since
we take all hypermultiplets to have unit charge, the coefficients
$k_{ij}$ must be rational in the compact case; then there is a basis
where all $k_{ij}$ and all hypermultiplet charges are integers.  Note
that Green's functions which are well-defined in the non-compact case
are identical in the compact case and are continuous functions of
$k$.}  In what follows the parameters $k_{ij}$ will be referred to as
Chern-Simons couplings, while the coefficients of the Maxwell terms
will be called gauge couplings, as before.  Note that the \nfour\
theories with BF couplings considered above form a subset of the set
of \nthree\ CS theories.

\nthree\ gauge theories have in general both Coulomb and Higgs
branches.  When $k_{ij}$ is nondegenerate, all vector multiplets
become massive and the Coulomb branch is lifted, while the Higgs
branch remains (if the number of hypermultiplets exceeds the number of
vector multiplets.)  Quantum corrections cannot lift this branch but,
unlike the \nfour\ case, they can modify its metric. The form of the
quantum corrections to the metric is tightly constrained by the
requirement that the metric be hyperk\"ahler.

Consider first \nthree\ SQED-1 with a CS coupling $k$ and infinite
bare gauge coupling.  In this case there is no moduli space: the
Coulomb branch is lifted because the CS term gives the vector
multiplet a topological mass, while the Higgs branch is lifted because
integrating out the vector multiplet produces a potential for the
hypermultiplet.  When $k\ra\infty$ the vector multiplet decouples, and
the theory becomes a theory of a free massless hypermultiplet, with
\nfour\ supersymmetry and $SU(2)_R\times SU(2)_N$ R-symmetry.  Since
the coefficient of the CS term is not renormalized 
\cite{csrenblasi,csrenkp2}, the theory with
$k\neq \infty$ is an exactly marginal deformation of the free
hypermultiplet, in analogy to the BF theory considered earlier.  One
can perform an ordinary Feynman diagram expansion in $1/k$, which by
power counting and supersymmetry is completely finite.  Since there
are no dimensionful parameters, there is no wave-function
renormalization of the hypermultiplet.  This also follows from the
fact that chiral gauge-invariant operators like $Q\Qt$ belong to short
representations of \nthree\ superconformal algebra, and their
dimension is determined entirely by their $SU(2)_D$ spin via
$\Delta=j_D$. The dimensions of nonchiral operators will generally depend
on $k$.

In the opposite limit $k\ra 0$ we obtain \nfour\ SQED-1 with infinite
gauge coupling, i.e.  CFT-1, which is mirror to the free
hypermultiplet found as $k\ra\infty$.  As in the BF case, we are led
to the conclusion that \nthree\ SQED-1 with CS coupling $k$ is dual to
\nthree\ SQED-1 with CS coupling $-1/k$.  (The inversion of the CS
coupling was previously argued, using branes in Type IIB string
theory, in~\cite{kooCS}.)  The generating functional of CFT-1 with CS
coupling $k$ is
\be{CSone}
Z_k[\UU]=\int \DD\VV\ {\rm Sdet}\left(\KK[\VV]\right)
\exp\left(ikS_{CS}(\VV,\VV)+iS_{GF}(\VV)+iS_{BF}(\UU,\VV)\right) \ . 
\ee
Upon using \Eref{master} and performing the Gaussian integral over $\VV$ 
this
becomes
\be{CStwo}
Z_{-1/k}\left[-{1\over k}\UU\right]\exp\left(-{i\over 
k}S_{CS}(\UU,\UU)\right)\ .
\ee
This shows that the connected correlators of $\VV$ at CS coupling $k$
and CS coupling $-1/k$ are related by a simple rescaling (and a shift
by a contact term for the two-point function).

Our discussion can be easily generalized to theories with more
multiplets.  Consider $N_f$ copies of CFT-1 and couple the vector
multiplets together as in \eref{CS}.  The theories at the origin of
moduli space make up a manifold of \nthree\ SCFTs parameterized by the
matrix $k$.  Mirror symmetry acts on this manifold by $k\ra -k^{-1}$.
For generic $k$ there is no moduli space; the only solution of
the classical D-flatness equations
\be{CSmoduli}
{1\over 2\pi}\sum_j k_{ij}\Phi_j^p  = Q^\dagger_i\sigma^p Q_i\equiv H_i^p,
\quad |\Phi_i|^2|H_i|^2=0
\ee
(no sum on $i$) is the trivial one, $\Phi=Q=\Qt=0$. In general, we may
search for solutions as follows.  The second set of equations in
\eref{CSmoduli} requires that we divide the indices $i$ into two sets,
which without loss of generality (through relabeling) we may take to
be $I=1,2,\cdots,n$ and $r=n+1,\cdots N_f$, and set
$H_I^p=0,\Phi_r^p=0$.  The equations
$$ k_{IJ}\Phi_J^p=0 $$
have nontrivial solutions if $k_{IJ}$ has zero determinant.  If this
is the case, then $\Phi_J^p= \sum_v c^p_v e_J^v$, where the $e_J^v$
are the zero modes of the minor $k_{IJ}$, and $c^p_v$ are three sets
of coefficients, $p=1,2,3$.  If we expand the photons $A^\mu_J$ as
$A^\mu_J=\sum_v b^\mu_v(x) e_J^v+ \ldots$, then the fields
$b^\mu_v(x)$ do not couple to themselves via CS terms, so their dual
scalars $\tau_v$ may be defined in the usual manner. The fields
$b^\mu(x)$ do couple to other photons, via BF-type terms.  As a result
of this the scalars $\tau_v$ transform additively under gauge
transformations of other photons.  Meanwhile, the other equations
$$ H_r^p = \frac{1}{2\pi} k_{rJ}\Phi_J^p $$ 
fix the expectation values of $Q_r$ up to a gauge transformation.  As
in the BF case, these equations can be interpreted as moment map
equations for a hyperk\"ahler quotient. When Maxwell terms are
present, the corresponding hyperk\"ahler metric is again of the
LR/LWY type~\cite{STdual,GR}.  An interesting issue is whether there
are quantum corrections to this metric.  The hyperk\"ahler property of
the metric and the presence of triholomorphic $U(1)$ isometries
(coming from the shift symmetries of the dual photons) ensure that the
quantum metric remains of the LR/LWY type.  However, these
considerations still allow the parameters of the metric to depend on
the elements of the matrix $k$ in an arbitrarily complicated manner.
We have not resolved this issue completely.

These results contain, as special cases, our results on \nfour\
theories with BF couplings and the mirror relations considered in
\cite{kooCS}.  Furthermore, the general \nthree\ CS theory can be
reduced to this example by linear field redefinitions and possible
addition of decoupled ($k=\infty$) vector and/or hypermultiplets.

It is interesting to note that duality of certain Chern-Simons
theories with respect to the inversion of the CS coupling has been
conjectured to underlie the structure of the phase diagram of quantum
Hall liquids; see for example~\cite{ShapWilcz,KLZ,LutRoss}.

\section{Piecewise Mirror Transformations}

Up to this point we have limited ourselves to discussing mirror
transformations applied to a theory as a whole, converting all
ordinary multiplets to twisted multiplets and vice versa.  However,
nothing prevents us from applying the mirror transform, as given in
\Eref{master}, to one hypermultiplet or twisted hypermultiplet at a
time.  We will call this operation a ``piecewise mirror transform.''
In general a theory with $p$ hypermultiplets and $q$ twisted
hypermultiplets will have $2^{p+q}$ piecewise-mirror descriptions.

To illustrate this we consider the simplest non-trivial example.  Take
$U(1)$ with two hypermultiplets, one of charge $1$ and one of charge
$q$.  The usual mirror transform converts its infrared conformal field
theory to that of twisted $U(1)$ with two twisted hypermultiplets of
charge $1$ and $-1/q$.  (Note the sign of a hypermultiplet charge can
be changed by a field redefinition.)  If instead we apply a piecewise
mirror transform to the hypermultiplet of charge $1$, using
\Eref{master}, we will find a theory with the following content: a
vector multiplet coupled to a twisted vector multiplet with a BF
coupling $k=1$, a hypermultiplet of charge $q$ coupled to the vector
multiplet, and a twisted hypermultiplet of charge $1$ coupled to the
twisted vector multiplet. Rescaling the vector multiplet, we
may set the hypermultiplet charge to $1$ and the BF coupling to
$k=1/q$.  Thus, $U(1)$ with hypermultiplets of charge $1,q$ is
piecewise-mirror to the BF theory in \Eref{BFaction} with coupling
$k=1/q$.

If instead we apply the piecewise mirror transform to the
hypermultiplet of charge $q$, we will similarly find a BF theory with
coupling $k=-q$.  This is the mirror of the previous BF theory. 
The following four theories are thus piecewise-mirror
\bea{fourmirror}\nonumber
& & \\ \nonumber
&{\rm \mbox{BF}}(\hat\QQ,\hat\VV,\VV,\QQ)[k=-q]& \\ \nonumber
& & \\ \nonumber
 {\rm \mbox{CFT-2}}(\QQ_1,\QQ_2,\VV)[q_1=1,q_2=q] & &
 {\rm \mbox{CFT-2}}(\hat\QQ_1,\hat\QQ_2,\hat\VV)[q_1=1,q_2=-q^{-1}] 
\\ \nonumber
& & \\ \nonumber
&{\rm \mbox{BF}}(\QQ,\VV,\hat\VV,\hat\QQ)[k=q^{-1}]& \\ \nonumber
\eea 
Note that the self-duality (up to a sign) of $U(1)$ with two
hypermultiplets of equal charge is equivalent to the self-duality (up
to a sign) of the BF theory with $k=1$.  

As a final comment, we note that the compactness of $U(1)$ requires
that the ratio of the charges of the hypermultiplets
be rational.  It follows from this, and from mirror
symmetry, that both $q$ and $k$ must be rational.  This
is consistent with the condition on $k$ that we mentioned earlier.

It is easy to apply the piecewise mirror transform to other models,
including the Chern-Simons theories of the previous section and the
non-conformal field theories of the next.

\section{Mirror symmetry away from the infrared limit}

In this section we give a field-theoretic interpretation of the
so-called ``magnetic coupling''\footnote{The term ``magnetic
coupling'' is an unfortunate misnomer, as the relation between this
interaction and the electric gauge interaction is not
electric-magnetic duality.  Mirror symmetry exchanges particles and
vortices, which couple (in the absence of a Chern-Simons coupling) to
different photons.}  \cite{kinsddd} and explain how the mirror
transform can be extended away from the infrared limit. The ``magnetic
coupling'' affects the metric on the Higgs branch, as will be reviewed
below, and is mirror to the gauge coupling.  However, its field theory
origin has not previously been determined.  As we will now show, it is
a Fermi-type coupling --- that which is induced between (twisted)
hypermultiplets by the exchange of a massive auxiliary (twisted)
vector multiplet to which they are minimally coupled.  We will refer
to the theory of a single hypermultiplet coupled to this massive
auxiliary vector multiplet as super-Fermi theory (SFT).

An indirect way to check that the gauge and super-Fermi couplings are
mirror, and that the super-Fermi coupling is indeed the constant term
in the metric on the Higgs branch, is to consider SQED-2 (with fields
$\VV,\QQ_1,\QQ_2$) and its mirror of the same form (with twisted
fields $\hat\VV,\hat\QQ_1,\hat\QQ_2$).  We will take the bare electric
and ``magnetic'' couplings to be infinite.

The Coulomb branch is parameterized by the $SU(2)_N$ triplet $\vec{\Phi}$
and the scalar $\tau$ which is the electromagnetic dual of the photon.  The 
metric is specified in terms of a harmonic function $G(\vec{\Phi})$ 
\be{metric}
ds^2 = G(\vec{\Phi}) (d\vec{\Phi}^2) + G^{-1}(\vec{\Phi}) (d\tau + 
\omega\cdot d\vec{\Phi})^2 \ ,
\ee
where $\nabla\times\omega(\vec{\Phi})=\nabla G(\vec{\Phi})$.
In the presence of a mass term $\vec{m}$ (a triplet of
$SU(2)_N$) for $\QQ_1$ and a mass term $-\vec{m}$ for $\QQ_2$ the 
function 
$G$ is given by
\be{Cmetrica}
G = {1\over |\vec{\Phi}-\vec{m}|}+{1\over |\vec{\Phi}+\vec{m}|} \ .
\ee
We may obtain SQED-1 by integrating out $\QQ_2$, i.e. by taking $m$ large 
while keeping $\vec{\phi}=\vec{\Phi}-\vec{m}$ fixed. In this limit we get 
\be{Cmetricb}
G \approx {1\over |\vec{\phi}|}+{1\over 2|\vec{m}|} \ .
\ee
The constant term in $G$ is the gauge coupling induced at one-loop by
integrating out the massive field $\QQ_2$; the one-loop integral leads
to a Maxwell term ${1\over 2m} S_{V}(\VV)$.  The low-energy theory is
SQED-1 with an effective coupling $g^2_{eff}=2m$.

In the mirror theory, the same branch appears as the Higgs branch,
which is parameterized by three fields $\vec{N}$, triplets of
$SU(2)_N$ which are bilinear in the twisted hypermultiplets, along
with a fourth scalar whose relation to the underlying fields is more
complex.  The metric on the Higgs branch similarly depends on a
harmonic function $\hat{G}$ of $\vec{N}$ and a possible
Fayet-Iliopoulos parameter $\vec{\xi}$ which is mirror to the mass
term $\vec{m}$:
\be{Hmetrica}
\hat{G} = {1\over |\vec{N}-\vec{\xi}|}+{1\over |\vec{N}+\vec{\xi}|} \ .
\ee
The mirror of taking $\vec{m}=\vec{\Phi}-\vec{\phi}$ is to take
$\vec{\xi}= \vec{N}-\vec{n}$.  For $|\vec{\xi}|\approx |\vec{N}|\gg
|\vec{n}|$, the field $\hat\QQ_2$ condenses and gives mass to
$\hat\VV$, leaving the field $\hat\QQ_1$ behind.  In the limit where
$\xi$ is large $\hat{G}$ becomes
\be{Hmetricb}
\hat{G} = {1\over |\vec{n}|}+{1\over 2|\vec{\xi}|} \ .
\ee
What is the interpretation of the constant term in $\hat{G}$?
It must be the coupling of the leading dimension-four operator
induced in this broken gauge theory --- which is obviously the
super-Fermi interaction for $\QQ_1$ induced by the massive photon.
\footnote{Note that string theory considerations also support this
claim.  A D3 brane of finite length $L$ which ends on two parallel NS5
(D5) branes contains as its lightest multiplets a massless \nfour\
$U(1)$ vector multiplet (hypermultiplet) and a massive hypermultiplet
(vector multiplet) of mass of order $\sim 1/L$ \cite{ahew}.  The gauge
coupling of the vector multiplet in the NS5 case is also of order
$1/L$, and so the gauge coupling in one theory is related to the mass
of a vector multiplet in its mirror.}

A more direct argument involves the computation of the metric in the
presence of the super-Fermi interaction.  To give a precise definition
of this interaction, let us multiply both sides of \Eref{master} by
$\exp(iS_V(\hat{\VV})/g^2)$ and integrate over $\hat{\VV}$. The left
hand side becomes the partition function of twisted \nfour\ SQED-1
with bare gauge coupling $g$, while the right-hand side corresponds to
\nfour\ SQED-1 with infinite bare gauge coupling coupled via a BF
term to a twisted vector multiplet $\hat\VV$. The action of the latter
theory is
\be{SauxA}
S = S_{H}(\QQ,\VV)+S_{BF}(\hat\VV,\VV)+{1\over g^2}S_{V}(\hat\VV) \ .
\ee
This is what we call the super-Fermi theory (SFT).
Since the action for $\hat\VV$ is quadratic it can be integrated
over, leaving the action $S = S_{H}(\QQ,\VV)+{g^2}S_{Vaux}(\VV)$
with
\be{SVaux}
S_{Vaux}(\VV) =-{1\over 4\pi^2} \int d^3x \int \ d^4\ta\ 
\left\{
{1\over 4}\Sigma{1\over \Box}\Sigma - \Phi^\dag {1\over \Box} \Phi 
\right\} \ 
,
\ee
which in the Landau gauge becomes an explicit mass term for $\VV$, 
\be{SVmass}
S_{Vaux}(\VV) =-{1\over 4\pi^2} \int d^3x \int d^4\ta\ 
\left\{V^2-\Phi^\dag {1\over \Box} \Phi \right\} .
\ee
Thus $\VV$ acts as an auxiliary field at the classical level.  After
integrating it out, we find by direct if tedious computation that the
action for $\QQ$ is that of a sigma-model with the Taub-NUT target
space. Moreover, the asymptotic radius of the circle parameterized by
$\tau$ agrees with that computed from the mirror SQED-1 theory.
(Another way of doing the same computation, using hyperk\"ahler
quotients, was explained in Section IV.)  The hyperk\"ahler property
of the metric ensures that there are no quantum corrections to this
result.

So far we showed the moduli space metrics of SQED-1 with finite gauge
coupling and twisted SFT agree, i.e. that the two theories are
equivalent in the extreme infrared everywhere on the moduli space. We
now claim that this equivalence is exact, so that \nfour\ SQED-1, in its
renormalization group flow from weak to strong coupling, is mirror to
twisted SFT at all energy scales.

This seems to be a very strong claim, as most known field theoretic
dualities have been established only in the infrared or for conformal
field theories.  However, if one has two well-defined exact
descriptions of an ultraviolet fixed point, then all perturbations of
this fixed point and the resulting renormalization group flows can be
described using the two sets of variables.  This is the case here.
The ultraviolet fixed point of which SQED-1 is a perturbation is a free
theory of a hypermultiplet and a vector multiplet which are not
coupled to one another.  This CFT has a mirror description as a copy
of twisted CFT-1 along with a vector multiplet to which it is not
coupled.  Consider the relevant perturbation given on one side by
coupling the vector multiplet to the flavor current of the
hypermultiplet, and on the other side by coupling the vector multiplet
to the global current $^*\hat F$ of the twisted CFT-1 via a BF term.  This
makes the first theory into SQED-1 with a weak gauge coupling and the
second into a theory of a twisted hypermultiplet coupled to an
auxiliary vector multiplet, which induces a large super-Fermi
coupling.  The gauge coupling in SQED-1 grows, and in the infrared the
theory becomes CFT-1.  The super-Fermi coupling in the mirror theory
shrinks, and in the infrared the twisted hypermultiplet becomes free.
To restate the claim, mirror symmetry implies
\be{flow}
\begin{array}{lcccc}
UV: & {\rm free} \ \QQ \ + \ {\rm free} \ \VV  & \Longleftrightarrow 
       & {\rm\ twisted\ \mbox{CFT-1}}(\hat\QQ,\hat\VV) + {\rm free}\ \VV 
\cr  
& & & \cr
\Downarrow & \Downarrow & & \Downarrow \cr \cr 
& & & \cr
{\rm flows \ to} \ \ \ \  
  & {\rm \mbox{SQED-1}}(\QQ,\VV) \ [{\rm coupling} \ g^2] 
&  \Longleftrightarrow 
  & {\rm twisted\  SFT}(\hat\QQ,\hat\VV,\VV) \ [{\rm coupling} \ {1/g^2}]\cr 
& & & \cr
\Downarrow& \Downarrow & & \Downarrow \cr 
& & & \cr
IR: & {\rm \mbox{CFT-1}}(\QQ,\VV)  & \Longleftrightarrow 
    & {\rm free\ twisted}\ \hat\QQ \cr
\end{array}
\ee

A corollary of this equivalence is that all correlation functions of
$\Sigma$ and $\Phi$ in SQED-1 must precisely agree with those of the
$U(1)$ current multiplet in the twisted SFT. This can be seen
explicitly from our master equation \Eref{master}. To this end
multiply both sides of \Eref{master} by
$$\exp\left\{{i\over g^2}S_{V}(\hat\VV)+iS_{BF}(\hat\VV,\VV')+
iS_{GF}(\hat\VV)\right\}\ $$
and integrate over $\hat\VV$. After performing a Gaussian integral over
$\hat\VV$ on the right-hand side and shifting the integration variables, one 
gets
\bea{SQEDSFT}
\int\ \DD\hat\VV\ 
&\exp\left\{{i\over g^2}S_{V}(\hat\VV)+
iS_{GF}(\hat\VV)+iS_{BF}(\hat\VV,\VV')\right\}
{\rm Sdet}\left(\KK[\hat \VV]\right)= \nonumber\\ 
& \ \ \int\ \DD\VV\ \exp\left\{{ig^2}S_{Vaux}(\VV)+iS_{GF}(\VV)\right\}
{\rm Sdet}\left(\KK[\VV-\VV']\right)\ .
\eea
Here the left-hand side is the generating functional for the
correlation functions of $\hat\Sigma$ and $\hat\Phi$ in (twisted)
SQED-1, while the right-hand side is the generating functional for the
correlators of the hypermultiplet's $U(1)$ flavor current in SFT.

An interesting implication of this result is that the perturbative
expansion of \Eref{SQEDSFT} in $g$ is the superrenormalizable SQED
expansion around a free theory, while the perturbation series in $1/g$
is the usual nonrenormalizable SFT expansion around a free theory.
The former expansion is finite, while the latter requires
renormalization and fails in the ultraviolet. We see that despite the
failure of the usual perturbative expansion in SFT, the theory still
has a perfectly well defined UV fixed point, as in the five and six
dimensional field theories considered first in
\cite{nsddddd,nsdddddd}.

It is also instructive to consider the current-current correlation
function in SQED.  For example, consider SQED-1 with a non-zero
Fayet-Iliopoulos term $\vec\xi$, which gives the hypermultiplet an
expectation value, $\vev{Q^\dagger \vec\sigma Q}=\vec\xi/(2\pi)$.  If
$|\xi|\gg g^2$, the photon is massive ($m_\VV^2\approx g^2\xi/\pi$) and
stable, and shows up as a single particle state in the two-point
function of $^*F$.  In addition there are much heavier semiclassical vortex
states, with $m^2_{\hat\QQ}=\xi^2$, which can be pair-produced by the
current.  Note that the vortex mass is protected by a BPS bound while that
of the photon is not.  As we reduce $|\xi|/g^2$, the photon and vortex
masses approach each other.  It is possible, for sufficiently small
$|\xi|$, that the photon becomes unstable and decays into vortices,
leaving no stable one-particle states in this channel.  Does this
occur?

For small $|\xi|$ the original variables are strongly coupled, so we
must use the mirror variables, which describe massive vortices of mass
$|\xi|$ weakly interacting via a short-distance potential. The
potential energy of a configuration of vortices is zero, but for a
configuration of both vortices and antivortices it is negative.  It is
known that two non-relativistic particles with an attractive
delta-function potential in two spatial dimensions have a single bound
state with exponentially small binding energy \cite{solvable}. We
therefore expect a single irreducible supermultiplet of stable
vortex-antivortex bound states. This supermultiplet is an ordinary
\nfour\ vector multiplet.  It therefore appears that there is a stable
massive vector multiplet in the theory for any value of $|\xi|/g^2$,
only merging into the continuum of vortex-antivortex states at
$\xi=0$.  We believe this is a new result that could not have been
derived without the identification of the magnetic coupling.

Let us find the binding energy of this bound state.  The coefficient
of the delta function in the low-energy non-relativistic theory is
logarithmically divergent.  To obtain a sensible result we must match
it to the coefficient of the Fermi interaction in the relativistic SFT
theory.  Supersymmetry ensures the relativistic Fermi interaction
receives only finite corrections, which are small if $|\xi|\ll g^2$.
Matching requires a cutoff, which should be at the scale of the
breakdown of the non-relativistic theory, that is, of order $|\xi|$.
Putting this together with the known result \cite{solvable}, we find
the binding energy is of order $-|\xi| e^{-g^2/2\pi|\xi|}$.

The results of this section can be easily extended to theories with
more flavors.  We mention one amusing example with a self-mirror
renormalization group trajectory.  Consider CFT-2, the infrared limit
of SQED-2, which is self-mirror.  The theory has two global symmetry
currents, a flavor current and a topological current exchanged under
mirror symmetry.  Using a vector multiplet $\VV$ and a twisted vector
multiplet $\hat\VV$, we may gauge both currents with equal
couplings. The resulting theory
\be{selfflow}
\int\ \DD\VV_0\ \DD\hat\VV\ \DD\VV\
e^{{i\over g^2}[S_{V}(\VV)+S_{V}(\hat\VV)]}
e^{iS_{BF}(\hat\VV,\VV_0)} \ 
{\rm Sdet}\left(\KK[\VV_0+\VV]\right)\ 
{\rm Sdet}\left(\KK[\VV_0-\VV]\right)\ ,
\ee
(here gauge fixing terms and couplings to background sources are
omitted for brevity) flows from CFT-2 plus free vector and twisted
vector multiplets in the ultraviolet to CFT-2 in the infrared.  The
flow can easily be seen, using \Eref{master}, to be self-mirror at
all scales.

\section{Vortex-creation operators}

Up to this point our discussion has been mostly concerned with the
action of mirror symmetry on conserved currents and their
superpartners. But if we want to make precise the statement that
mirror symmetry exchanges particles and vortices \cite{ntwovort}, we
need to understand vortex-creation operators in SQED.

The gauge-invariant vortex-creation operators are associated with some
of the most poorly understood aspects of mirror symmetry.  Mirror
symmetry unambiguously implies that such operators must be present in
the CFTs that are found at the origin of moduli space.  However, the
only hint as to how to define them is found far along the moduli space
of the Coulomb branch, where all of the charged matter is massive.
There, the low-energy theory involves only the vector multiplet, and
one may safely replace each photon with its dual scalar $\tau$. Vortex
creation operators are known to be proportional to $e^{i\tau}$.  From
mirror symmetry we know that some of the vortex-creation operators are
chiral, in the \ntwo\ sense, and so, if the real scalar $\phi$ in the 
\ntwo\ vector multiplet has an expectation value, a natural form for a
vortex-creation operator is $e^{(i\tau+\phi/g^2)}$, where $g^2$ is
the low-energy effective gauge coupling.  However, a number of
puzzles surround this choice.  How are these operators to be continued
to the origin of moduli space, where there is a CFT involving massless
charged matter which prevents naive definition of $\tau$?  What is to
be done about the paradox that $\phi$ remains dimensionful in the CFT
but no scale $g^2$ remains in the theory?
Assuming these problems are resolved, how does the operator obtain its
correct conformal dimension?  How does it acquire its abelian global
charges?  For \nfour\ SQED, where this operator should be part of an
$SU(2)_N$ multiplet, what are the other operators in the multiplet and
how is the nonabelian global symmetry realized?  For those cases where
there are hidden flavor symmetries \cite{kinsddd,ntwovort} which must 
act on
these operators, how do those symmetries appear?

We will now attempt to provide answers to some of these questions.  To
set the stage, let us recall how to construct operators with nonzero
vortex charge in an arbitrary abelian gauge theory in three dimensions
(formally).  Consider a $U(1)$ gauge field coupled to massless
matter. Formally integrating over matter fields, we get a nonlocal
effective action for the gauge field $A^\mu$.  Gauge invariance tells
us that it can be regarded as a functional $S_{eff}(F)$ of the field
strength $F=dA$. Let us change variables in the path integral from
$A^\mu$ to $F^\mn$; since $F^\mn$ satisfies the constraint $dF=0$, we
must implement it using a Lagrange multiplier $\tau$:
\be{deftau}
\int\ \DD A^\mu\  \delta(\del_\mu A^\mu) 
e^{iS_{eff}\{F(A)\}}
\propto
\int\ \DD F^\mn\  \DD\tau\
\exp\left(iS_{eff}(F)+{i\over 2\pi}\int d^3x\ \tau dF\right)\ .
\ee
Note $^*F$ has dimension 2 (as demanded of a conserved current by the
conformal algebra) so $\tau$ is dimensionless and, in analogy to a
free boson in two dimensions, can be exponentiated.  It follows from
\eref{deftau} that $\tau$ is canonically conjugate to $\eps^{ij}
F_{ij}$, so that the symmetry transformation generated by the current
$^*F$ acts additively on $\tau$ and multiplicatively on $e^{in\tau}$.
Our normalization is such that $e^{i\tau}$ carries a unit of this
charge, the integrated magnetic flux.  

Since a vortex worldline carries magnetic flux, any operator which
creates a vortex must be proportional to $e^{i\tau}$ \cite{ntwovort}.
To see this, consider the correlation function of two such operators
\be{twosigma}
\int\ \DD F^\mn   \DD\tau
\exp\left\{iS_{eff}(F)+{i\over 2\pi}\int d^3x 
 \tau dF\right\}
e^{ i \tau(x)} e^{- i \tau(y)}\ .
\ee
The integration over $\tau$ gives a factor of $\delta[dF -
2\pi\delta(x)+2\pi\delta(y)]$.  Thus, the Bianchi identity is violated
by two pointlike sources of magnetic flux --- pointlike nondynamical
Dirac monopoles, which are instantons in three dimensions. On the
Higgs branch, where flux is confined into particle-like vortex
solitons, these pointlike instantons will indeed be sources for these
solitons.

A SUSY-covariant extension of this procedure can be constructed
following \cite{STdual}, where it was shown how to dualize an \ntwo\ vector
multiplet to a chiral multiplet on the Coulomb branch of the moduli space.  
The superspace
effective action for the $U(1)$ vector multiplet $V$ is regarded as a
functional of $\Sigma=iD\bar{D} V$.  $\Sigma$ satisfies supersymmetric
Bianchi identities $D^2\Sigma=0={\bar{D}}^2\Sigma$.  If we impose this
constraint explicitly by introducing a Lagrange multiplier chiral
superfield $T$ which couples to $\Sigma$ via $\int d^3x d^4\theta\
\Sigma(T+T^\dagger)$, then we may replace integration over $V$ by
integration over an unconstrained real superfield $\Sigma$. The
partition function for an \ntwo\ theory takes the form
\be{partntwo}
\int\ \DD \Sigma \DD T 
\exp\left(iS_{eff}(\Sigma)+{i\over 4\pi}\int d^3x d^4\theta\ \Sigma 
(T+T^\dagger)\right)\ .
\ee
The normalization in \eref{partntwo} is such that the imaginary part
of the lowest component of $T$ is $\tau$, so $e^T$ has vortex charge
$+1$.

Consider now \ntwo\ or \nfour\ SQED-$N_f$.  The mirror of \ntwo\
SQED-$N_f$ differs from that of \nfour\ SQED-$N_f$, which was
described in Sec.~II, only by the presence of an extra neutral chiral
superfield which couples to $\sum_p\hat{Q}_p\hat{\Qt}_p$
\cite{ntwobrane,ntwovort}.  In both cases the mirror $U(1)^{N_f-1}$ gauge
theory has \ntwo\ chiral primary operators $V_+=\hat{Q}_1\ldots
\hat{Q}_{N_f}$ and $V_-=\hat{\Qt}_1\ldots\hat{\Qt}_{N_f}$. Their
vortex charges are $+1$ and $-1$, respectively. They live in a short
representation of the \ntwo\ or \nfour\ superconformal algebra, and
therefore their dimensions are related to their R-charges.  In \nfour\
SQED-$N_f$ the dimensions are fixed to be the canonical dimension,
$N_f/2$.  In \ntwo\ SQED-$N_f$ the dimensions are not
known, since the theory has a one-parameter family of R-currents from
which it is not clear how to select the relevant one, but in the large
$N_f$ limit the R-charges and dimensions can be determined using
mirror symmetry, as we will now explain.

As is well-known, non-supersymmetric QED is completely solvable in
this limit \cite{Appelquist:1981}. The effective action for the photon
given by integrating out $N_f$ massless electrons is simply
$\Lgr_{eff} \propto N_f F^\mn[-\Box]^{-1/2}F_\mn$ plus higher orders
in the field strength.  As always, in the large $N_f$ limit all
scattering is suppressed and the theory becomes Gaussian; since the
photon propagator is nonstandard, it is known as a ``generalized free
field.''\footnote{In position space the photon propagator is
proportional to $1/x^2$.  It was pointed out long ago that this is the
same as the four-dimensional photon propagator projected down onto a
three-dimensional hyperplane.  We may observe that
it is also the projection onto the boundary of four-dimensional
Anti-de Sitter space of a photon propagating on that space.  To be
more precise, for a background gauge field coupled to the
three-dimensional current $^*F$, the induced propagator in
three-dimensions will also be $[-\Box]^{-1/2}\sim 1/x^2$, as though it
were a free massless vector field on $AdS_4$.  This is not to suggest
large $N_f$ (S)QED has a (super)gravity dual; the form of the propagator is
fixed by conformal invariance alone.}  Similarly, for \nfour\ SQED in
the large $N_f$ limit one gets
\bea{largenf}
Z_{N_f}[\hat\VV]& = &  
\int\ \DD\VV\  e^{iS_{BF}(\hat\VV,\VV)+iS_{GF}(\VV)} 
\ \left[{\rm Sdet}\left(\KK[\VV]\right)\right]^{N_f}\\ \nonumber
&\approx &\int\ \DD\VV\ e^{iS_{BF}(\hat\VV,\VV)+iS_{GF}(\VV)}
\exp\left\{{i N_f\over 16}\int d^3 x\int d^4\theta\ 
\left(\Sigma[-\Box]^{-1/2}\Sigma
-4\Phi^\dagger[-\Box]^{-1/2}\Phi\right)\right\}   
\eea 
with an analogous expression for \ntwo\ SQED.  Thus the vector
multiplet is described by a supersymmetric generalized free field. The
dimensions of matter fields $Q,\Qt$ in SQED-$N_f$ are canonical up to
corrections of order $1/N_f$~\cite{Appelquist:1981}, so the mesons
$\Qt_p Q_p$ and their mirrors $S_p$~\cite{ntwobrane,ntwovort} have
dimension $1$.  It follows that the dimensions of $\hat{Q}_p$ and
$\hat{\Qt}_p$, which couple to $S_p$ in the superpotential
$W=S_p\hat{Q}_p\hat{\Qt}_p$, are canonical, so $V_+$ and $V_-$ both
have dimension $N_f/2$.

Consider now operators $e^T$ and $e^{-T}$ in \ntwo\ or \nfour\
SQED-$N_f$, where $T$ is the dual photon superfield defined above.
These operators are (naively) chiral and have vortex charge $+1$ and
$-1$, and the operators $V_+$ and $V_-$ should therefore be
proportional to them.  The dimensions of $e^{\pm T}$ match those of
$V_{\pm}$ in the large $N_f$ limit, as we now show by computing the
two-point function of the lowest component of $e^{\pm T}$.  Using
\eref{largenf} and performing the Gaussian integral over $\Sigma$ we
find:
\be{taucorr}
\vev{e^{T(x)} e^{T^\dagger(y)}}\sim \int \DD T
\exp\left(T(x)+T^\dagger(y)- {i\over 2\pi^2 N_f} \int d^3z d^3z'
T(z)[-\Box]^{3/2} T^\dagger(z')\right)\ . 
\ee 
$T$ is dimensionless and its propagator $[-\Box]^{-3/2}$ is
logarithmic in position space, so the operator $e^{T}$ has a
well-defined dimension.  Performing the Gaussian integral over $T$ in
\eref{taucorr} we find that $e^{T}$ has dimension $N_f/2$.

In summary, we have clarified several issues.  The field $\tau$
can still be defined at the origin of the moduli space without
difficulty, as long as one first integrates out the massless charged
matter and re-expresses the resulting non-local action $S_{eff}$ using
$F_\mn$.  The complex scalar which is exponentiated is 
$$
T|_{\theta=0} = i\tau -{1\over 8\pi}{\delta S_{eff}\over\delta
\Sigma}\Big|_{\theta=0}\ ,$$ 
a non-local expression which nonetheless agrees with expectations far
along the Coulomb branch.  With proper normalization, the dimensions
of the vortex operators $e^{\pm T}$ have been shown to match those of
$V_{\pm}$ in the large $N_f$ limit, where $S_{eff}$ can be computed.

However, this is not the whole story.  Apart from their vortex charge,
the operators $V_{\pm}$ carry non-zero and equal abelian R-charges.
This is connected with the fact that in \nfour\ SQED-$N_f$ there is an
operator relation of the form $V_+ V_-\sim \Phi^{N_f}$.  It is
impossible for $e^T$ and $e^{-T}$ to satisfy these constraints.  Even
more confusing is the fact that in \nfour\ SQED-$N_f$ the operators
$V_+$ and $V_-^{\dagger}$ actually belong to a spin-$N_f/2$ multiplet
of $SU(2)_N$.

To resolve these issues, care should be taken in
the definition of the operators $e^{\pm T}$.  As in two dimensions,
the presence of a logarithmic propagator implies the need for an infrared
regulator, which should be supersymmetry-preserving.  This regulator
may carry global symmetry charges, which might resolve some of the
remaining puzzles.  It is also possible that one must account for
fermionic zero-modes of the pointlike Dirac monopoles that $e^{\pm T}$
are intended to represent.

As a last comment, we note that one of the most important unsolved
problems in mirror symmetry is the mapping of the full nonabelian
flavor symmetries.  \nfour\ SQED-$N_f$ has an $SU(N_f)$
flavor symmetry, but in the mirror description only the diagonal
generators are visible classically, with the rest emerging through
quantum effects \cite{kinsddd}.  This can be seen from the fact that
operators in nontrivial representations of $SU(N_f)$ appear in the
mirror theory as a combination of operators built from fundamental
fields with other operators built from vortex-creation operators
\cite{ntwovort}.  A proper definition of the vortex-creation
operators is a prerequisite for an understanding of the hidden
symmetries.

\section{Outlook}

We have found an elegant formula, Eq.~\eref{master}, which summarizes
many known results of mirror symmetry.  The formula states that the
superdeterminant of the \nfour\ supersymmetric Laplacian on
three-dimensional Minkowski space is its own generalized Fourier
transform.  We have used it to find new superconformal field theories
with exactly marginal couplings, on which mirror symmetry acts as
strong-weak coupling duality.  We have established mirror relations
between non-conformal theories which are valid at all energy scales.
Finally, we have made some progress toward understanding how to define
the vortex-creation operators which appear in these theories.
However, many questions remain.  We do not have the precise change of
variables underlying mirror symmetry, which requires a clearer
understanding of vortex operators. We have no proof of our formula
from first principles, and see no hint of a reformulation of the
theory in which it would be manifest.  Lastly, we have no idea how to
generalize it to non-abelian gauge theories.  We hope that future
research will overcome these obstacles to a more profound
understanding of duality.

\bigskip

We thank M. Berkooz, S.J.~Gates, K. Intriligator,
N. Seiberg, F. Wilczek and E. Witten for discussions.  The work
of A.K. was supported by Department of Energy grant
DE-FG02-90ER40542; that of M.J.S. was supported in
part by National Science Foundation grant NSF PHY-9513835 and
by the W.M.~Keck Foundation.
   
%  \nocite{*}                %this uses *everything* in the .bib file
   \bibliography{FT}        %or whatever your .bib file is
\bibliographystyle{utphys}   %if you use utphys.bst

\end{document}